\documentclass[aps,prl,twocolumn,showpacs,superscriptaddress,amsmath,amssymb]{revtex4}

\usepackage{graphicx}
\usepackage{color}

\definecolor{Red}{rgb}{1.0,0.0,0.0}

\begin{document}

\title{Fracturing highly disordered materials}

\author{A. A. Moreira} \affiliation{Departamento de F\'{\i}sica,
  Universidade Federal do Cear\'a, 60451-970 Fortaleza, Cear\'a,
  Brazil}

\author{C. L. N. Oliveira} \affiliation{Departamento de F\'{\i}sica,
  Universidade Federal do Cear\'a, 60451-970 Fortaleza, Cear\'a,
  Brazil}

\author{A. Hansen} \affiliation{Department of Physics, Norwegian
  University of Science and Technology, N-7491 Trondheim, Norway}

\author{N. A. M. Ara\'{u}jo} \affiliation{Computational Physics for
  Engineering Materials, IfB, ETH Z\"{u}rich, 8093 Z\"{u}rich,
  Switzerland}

\author{H. J. Herrmann} \affiliation{Departamento de F\'{\i}sica,
  Universidade Federal do Cear\'a, 60451-970 Fortaleza, Cear\'a,
  Brazil} \affiliation{Computational Physics for Engineering
  Materials, IfB, ETH Z\"{u}rich, 8093 Z\"{u}rich, Switzerland}

\author{J. S. Andrade Jr.}  \email{soares@fisica.ufc.br}
\affiliation{Departamento de F\'{\i}sica, Universidade Federal do
  Cear\'a, 60451-970 Fortaleza, Cear\'a, Brazil}

\date{\today}
\pacs{64.60.ah, 64.60.al, 89.75.Da}

\begin{abstract}  
  We investigate the role of disorder on the fracturing process of
  heterogeneous materials by means of a two-dimensional fuse network
  model. Our results in the extreme disorder limit reveal that the
  backbone of the fracture at collapse, namely the subset of the
  largest fracture that effectively halts the global current, has a
  fractal dimension of $1.22 \pm 0.01$. This exponent value is
  compatible with the universality class of several other physical
  models, including optimal paths under strong disorder, disordered
  polymers, watersheds and optimal path cracks on uncorrelated
  substrates, hulls of explosive percolation clusters, and strands of
  invasion percolation fronts. Moreover, we find that the fractal
  dimension of the largest fracture under extreme disorder, $d_f=1.86
  \pm 0.01$, is outside the statistical error bar of standard
  percolation. This discrepancy is due to the appearance of trapped
  regions or cavities of all sizes that remain intact till the entire
  collapse of the fuse network, but are always accessible in the case
  of standard percolation. Finally, we quantify the role of disorder
  on the structure of the largest cluster, as well as on the backbone
  of the fracture, in terms of a distinctive transition from weak to
  strong disorder characterized by a new crossover exponent.
\end{abstract}

\maketitle

The brittle fracture of heterogeneous systems still represents a major
challenge from both scientific and technological points of view. It
has been the subject of intense scientific research in physics,
material science, mechanical and metallurgical engineering
\cite{Marder2010,Kanninen1985,Sahimi1998,Alava2006}. The amount of
small cracks and the shape of the largest fracture potentially
determine whether or not the system still supports external loads and
how catastrophic will be the rupture. Previous studies have shown that
the degree of disorder in the material rules the transition between
abrupt and gradual ruptures, i.e., the more heterogeneous is the
material, more warnings one gets before the system collapses
\cite{Herrmann1988}.  However, the nature of this transition, as well
as the behavior of the fracturing system in the limit of strong
disorder \cite{Malakhovsky2007,Hao2012}, remain as important open
questions.  Evidently, to model fracturing formation, it is
fundamental to determine how stress is redistributed over the system
while cracks appear, grow, and merge. Although conceptually simple,
the fuse network model is a good candidate for dealing with such
problem, since it can clearly capture the essential features of the
involved physical phenomena. In this model, resistors are used to
mimic springs, in an approximate description for elasticity, where
vector and tensor fields describing fracture mechanics are replaced by
a scalar field representing the local strain
\cite{Hansen1991,Arcangelis1989,Gilabert1987,Zapperi2008}.

The purpose of this Letter is to investigate the role of disorder on
the scaling properties of the fuse network model at the critical
collapse condition. We first study in detail the limiting case of
extreme disorder. This is performed by measuring, through a purely
geometrical technique, the fractal dimension of three different
structures, namely the {\it backbone} of the fracture of broken bonds
that halts current through the lattice, the {\it largest fracture}
formed by the broken bonds (backbone plus dangling ends attached to
it), and the {\it total} network of all broken bonds (largest cluster
plus smaller clusters not attached to it).  We then show how the
self-similar behavior of the resulting fracture topology crosses over
from one regime to another as we move from weak to strong disorder.
\begin{figure}[t]
  \includegraphics[width=8cm]{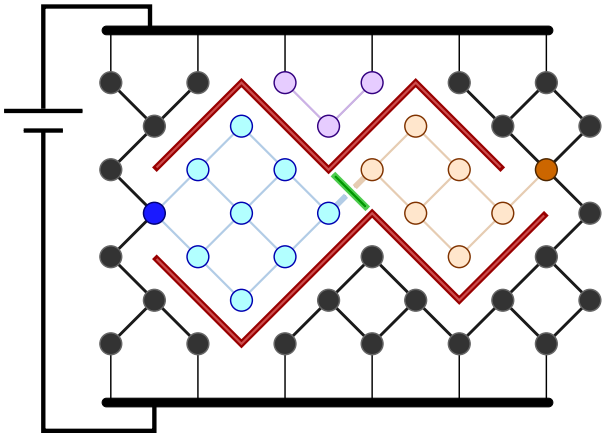}
  \caption{(Color online) The fuse network under extreme disorder. In
    each step, a random bond belonging to the conducting part of the
    network is chosen to burn. Here the burnt bonds were removed and
    the red lines placed instead, but in the complementary lattice.
    The red lines represent cracks in the system and the green line
    close to the middle of the panel corresponds to a bond that will
    merge two large cracks, forming two cavities in the lattice (light
    blue and yellow dots).  Since the current can not go through, none
    of the bonds inside the cavities (light blue and yellow lines) can
    burn in the next steps.  The dark blue and brown dots depict the
    limiting boundaries of each cavity. Note that the resulting
    fracture forms no loops. The purple dots in the middle top of the
    lattice correspond to sites connected to only one of the poles,
    therefore they do not participate in the transport, and the
    (purple) bonds connected to them do not burn either.}
  \label{f.fuse}
\end{figure}

Let us start by describing the fuse network model
\cite{Arcangelis1985,Herrmann1988}. In the traditional version of this
model, the electric potential in a resistor network should provide a
simplified description of the local strain in a fracturing system. A
crack forms when the stress or, correspondingly, the electric current
over a given resistor, surpasses a certain threshold.  Therefore, our
system is a lattice where each bond is a resistor with a given
conductance and fusing threshold value. For simplicity, we consider
here the case in which the conductance is the same for all bonds,
however, we expect similar results with a varying conductance
\cite{footnote1}. We model the heterogeneity by assigning to each bond
$i$ a threshold given by $\tau_i=10^{\beta R_i}$, where $R_i$ is a
random number uniformly distributed in the interval $-1<R_i<1$.
Therefore, the distribution of thresholds is hyperbolic,
$P_\tau(\tau)\sim\tau^{-1}$, with upper and lower bounds given by
$10^\beta$ and $10^{-\beta}$, respectively.

Generally speaking, we consider that a potential difference is applied
between two opposite sides of a resistor network, and solve
Kirchhoff's law to determine the current passing through each bond
\cite{HSL}.  Below the threshold value, each bond conducts according
to Ohm's law, but once the current $I_{i}$ at a given bond $i$ reaches
the threshold $\tau_{i}$, the bond burns (or breaks, in the mechanical
terminology) and becomes an insulator. In this way, the largest
current-threshold ratio, $max(I_{i}/\tau_{i})$, determines the next
bond to be burnt. Here we assume that the potential starts from zero
and raises slowly, allowing only one bond to burn at each step.
Pathological defects in the system are avoided by using a tilted
square lattice, therefore in the first step all currents are the same,
and the bond with the smaller threshold is the first to burn. In the
following steps, inhomogeneities are gradually introduced in the
system due to the burnt bonds, so that local threshold and current
values should determine which one burns next.

We initially focus on the case of a fuse network under extreme
disorder, $\beta\to\infty$, where thresholds are distributed over many
orders of magnitude. In this limit, for all practical purposes, one
can assume that the smaller (larger than one) ratio between the
thresholds of any two bonds in the lattice is larger than the largest
ratio of the currents of any two bonds that constitute the conducting
part of the network. In this regime, any variability in the currents
becomes irrelevant so that the next bond to burn should be the one
with the smaller threshold among those that participate in transport.
Since the thresholds are randomly distributed over the lattice, this
is entirely equivalent to a process in which a random bond with finite
current, $I>0$, is chosen to burn at each step.  As the bonds burn,
however, they may form a cavity that is connected to the rest of the
lattice at a single node, as represented in Fig.~\ref{f.fuse}. Because
all nodes inside a cavity are equipotential, their connecting bonds
are current-free and can not burn, regardless of their thresholds.
\begin{figure}[t]
  \includegraphics[width=8.5cm]{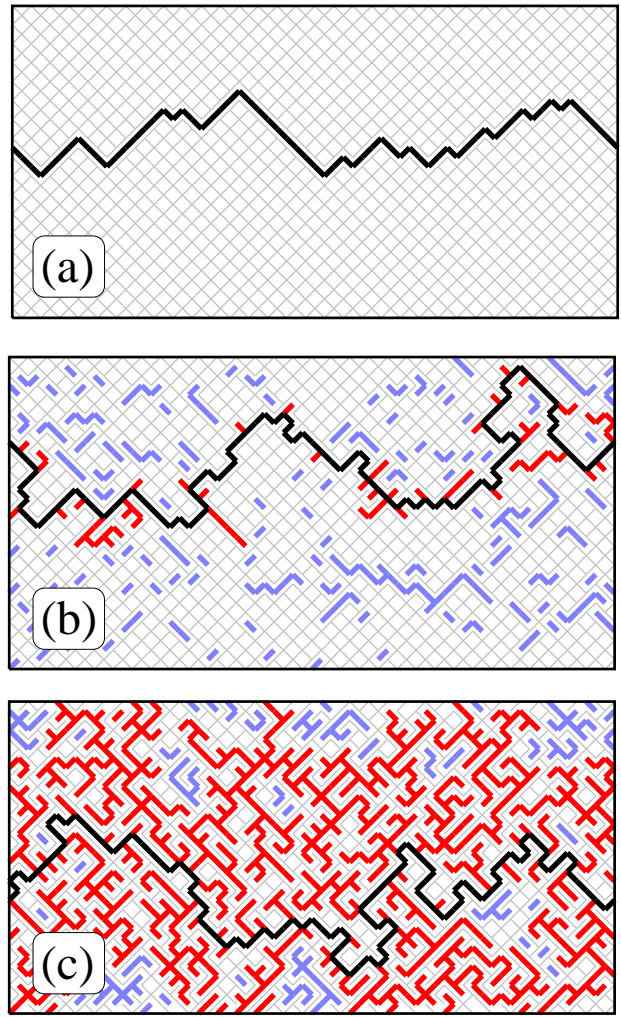}
  \caption{(Color online) Typical realizations of the fuse network at
    the point of disconnection ($L=32$). In (a) $\beta=0.01$ (weak
    disorder); in (b) $\beta=1$ (intermediate disorder); and in (c)
    $\beta=100$ (strong disorder, for this lattice size).  As
    represented in Fig.~\ref{f.fuse}, the fractures occupy the` places
    of the burnt bonds, but in the complementary lattice. The black
    lines show the {\it backbone} with mass $M_{b}$, namely the number
    of burnt bonds forming the chain that effectively disconnects the
    bottom from the top of system. The set of all cracks connected to
    form the {\it largest fracture}, namely the backbone and its
    dangling ends (black and red lines, respectively) have mass
    $M_{f}$, while $M_{t}$ denotes the {\it{total}} mass, i.e., the
    mass of the largest fracture plus all the other cracks (light blue
    lines) in the system.}
  \label{f.struc}
\end{figure}

In this way, the extreme disorder limit of the fuse network model can
be viewed as a modified percolation problem. In the standard
percolation, bonds are randomly and sequentially occupied, while in
the fuse network under extreme disorder, one only burns (occupies)
bonds that belong to the conducting backbone of the lattice. As shown
in Fig.~\ref{f.fuse}, once a cavity is formed, the bonds inside remain
unoccupied and, as a consequence, the clusters of occupied bonds in
the complementary lattice do not form loops. Under this reasoning, the
extreme disorder limit of the fuse network model becomes a purely
geometrical problem. Simulations performed with fuse networks at
$\beta=10^8$ confirm that these two models are identical. Such a
geometrical approach for the extreme disorder case greatly reduces the
computational demand of the problem, therefore enabling us to simulate
the fracturing process for networks with linear size $L$ going from
$16$ to $2048$, and using at least $1000$ samples for the largest
size. As shown in Fig.~\ref{f.struc}, we stop each realization when
the two sides, where the potential difference is applied, become
disconnected, i.e., no current can pass through the system.

\begin{figure}[t]
  \includegraphics[width=8cm]{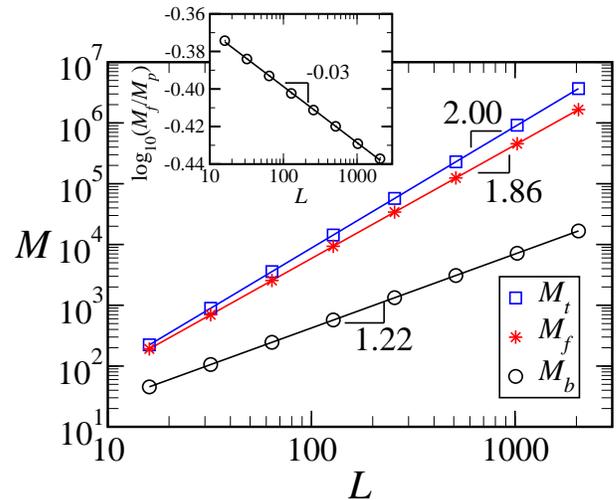}
  \caption{(Color online) Size dependence results obtained for the
    fuse network model under extreme disorder. The total number of
    burnt bonds in the system scales as, $M_{t}\sim{L^{d_{t}}}$, with
    $d_{t}=2.00\pm{0.01}$, indicating that the cracks appear
    homogeneously in the 2D lattice. The largest fracture for this
    model grows as $M_{f}\sim{L^{d_{f}}}$, with $d_{f}=1.86\pm{0.01}$,
    suggesting a different universality class for this structure than
    that of the spanning cluster in standard percolation
    \cite{Stauffer1992}.  Finally, the mass of the backbone of the
    largest fracture scales as $M_b\sim{L^{d_b}}$, with
    $d_b=1.22\pm{0.01}$. The results shown in the inset confirm that
    the ratio between the masses of the largest clusters in the fuse
    and in the percolation models scales as,
    $M_{f}/M_{p}\sim{L^{d_{f}-d_{p}}}$, with
    $d_{f}-d_{p}\approx{-0.03}$.}
  \label{f.strong}
\end{figure}

In Fig.~\ref{f.strong} we show that the mass of the largest fracture,
$M_{f}$, its backbone mass $M_{b}$, as well as the set of all broken
bonds, $M_{t}$, grow with the system size as power laws. First, the
total mass of burnt bonds scales with the linear size as,
$M_{t}\sim{L^{d_t}}$, with $d_t=2.00\pm{0.01}$, suggesting that the bonds
burn homogeneously through the lattice. The backbone grows as
$M_{b}\sim{L^{d_b}}$, with $d_b=1.22\pm{0.01}$, which is statistically
identical to the exponent obtained for optimal paths under strong
disorder \cite{Andrade2009}, disordered polymers \cite{Cieplak1994},
watersheds \cite{Fehr2009} and optimal path cracks on uncorrelated
substrates \cite{Oliveira2011}, the hulls of explosive percolation
clusters \cite{Araujo2010}, and the strands of invasion percolation
fronts \cite{Cieplak1996}. For the largest fracture, however, we
obtain $M_{f}\sim{L^{d_f}}$, with $d_f=1.86\pm{0.01}$, which is
different from the fractal dimension of the largest cluster in 2D
percolation, $d_p=1.8958$ \cite{Stauffer1992,Roux1988}.

In the limit of extreme disorder, the fracture backbone in the fuse
model is identical to the one of loopless percolation
\cite{Porto1997}.  This can be seen by considering that the burning of
fuses of a specific configuration of thresholds due to the extreme
disorder just follows the sequence of their inverse rank, except if
the fuse would close a loop or lie inside a nearly closed loop.  In
parallel, one can construct a configuration of loopless percolation by
assigning an occupation rank to each site of the lattice identical to
the ranking given by the thresholds (and obviously not occupying a
site which would close a loop). The spanning cluster of this
percolation configuration then consists of the fracture of the fuse
model and sites inside nearly closed loops that do not contribute to
the backbone.  This shows that the bonds forming the backbone should
be exactly the same in both models. It was previously observed that
the backbone of loopless percolation has the same fractal dimension as
the optimum path in strong disorder \cite{Schrenk2012}, therefore
giving support to this argument.

The cavities can not change the fractal dimension of the backbone but
may very well change the dimension of the largest fracture. The
largest fracture comprises the backbone and the dangling ends attached
to it.  Once a cavity is formed in the largest fracture, it precludes
the growing and attaching of other dangling ends inside it. As a
consequence, the largest fracture in the extreme disorder fuse model
is certainly smaller than the largest cluster in standard
percolation. Since we obtain, for the former case, a fractal dimension
smaller than the expected for standard percolation, we conclude
that cavities form at every scale, and the difference between the two
models becomes increasingly relevant as the system size grows.

To further test our hypothesis, we performed simulations with a
pinpoint algorithm that builds simultaneously percolation clusters and
fuse network fractures under extreme disorder.  Precisely, at each
step, we randomly choose a bond in the lattice and, if this bond is
part of the conducting backbone, it is occupied in the percolation
lattice as well as burnt in the fuse network. If this bond is inside
one of the cavities, however, it is occupied only in the percolation
lattice and not in the fuse network. Once a cluster forms that crosses
the system, the percolation condition is achieved in both models
simultaneously. If the fractal dimensions are in fact different, the
ratio between the masses of the largest clusters, $M_{f}/M_{p}$,
where $M_{p}$ is the mass of the spanning cluster in percolation,
should vary with the system size as a power law, with an exponent
given by $d_{f}-d_{p}\approx{-0.03}$. The obtained results shown in the
inset of Fig.~\ref{f.strong} support our conjecture that the largest
fracture of the fuse network under extreme disorder corresponds to a
subset of the spanning cluster in percolation. Moreover, this new
object is also self-similar, but with a slightly smaller fractal
dimension.
\begin{figure}[t]
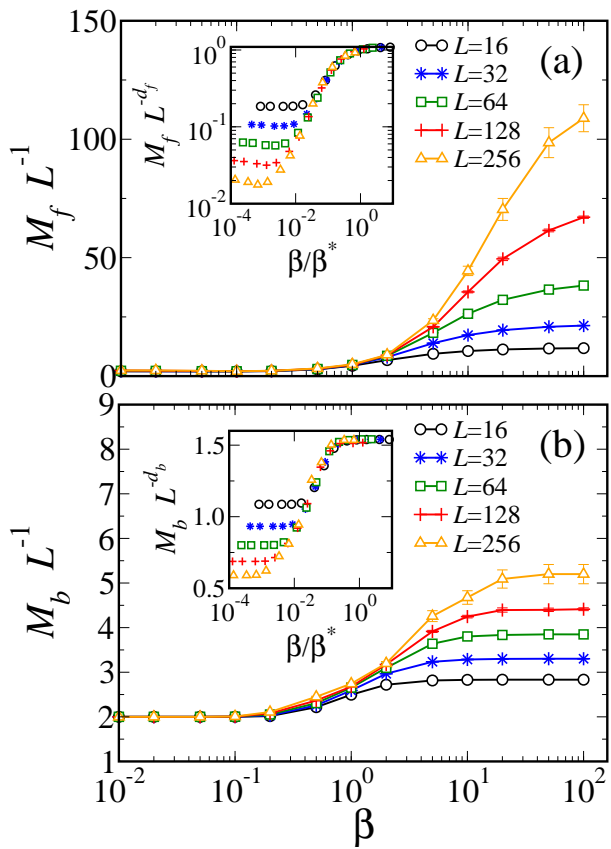

  \includegraphics*[width=8cm]{fig4a.eps}
  \includegraphics*[width=8cm]{fig4b.eps}
  \caption{(Color online) The transition from weak to strong disorder.
    We show the mass of the {\it largest fracture} (a) and {\it
      backbone} (b) as function of the disorder parameter $\beta$. In
    the week disorder regime, $\beta<\beta^*$, we have a linear growth
    for both backbone and largest fracture. As $\beta$ increases the
    masses cross over for the strong disorder regime where a
    super-linear growth is observed. The exponents $d_f$ and $d_b$,
    controlling the scaling in the strong regime, are the same as in
    Fig~(\ref{f.strong}). The value $\beta^*$ determining the onset of
    the crossover scales with the size of the system as
    $\beta^*\sim{L^{0.9}}$. }
  \label{f.beta}
\end{figure}

It is important to mention that some variants of the percolation model
previously investigated also depart from the universality class of
standard percolation. For example, in invasion percolation with
trapping \cite{Wilkinson1983}, when a loop is formed in the growing
cluster, no site or bond inside this loop can be occupied,
constituting a trapped region. These trapped regions are somehow
similar to our cavities, only that cavities are sections of the
lattice outside the conducting backbone of unoccupied bonds, while
trapped regions are sections outside the infinite cluster of
unoccupied bonds. In this case, statistically relevant deviations have
also be detected from the fractal dimension of the spanning cluster in
standard percolation \cite{Roux89,Sheppard99}.

Next we investigate how the behavior of the fuse network model crosses
over from weak to strong disorder by gradually increasing the value of
the parameter $\beta$. Local currents are computed by applying
Kirchhoff's law to each site of the network at each burning step, and
solving the resulting system of linear algebraic equations. In weak
disorder, most of the burnt bonds belong to the backbone of the
fracture that grows linearly with the system size, $d_b=d_f=1$. As
$\beta$ increases, the fractal dimensions obtained in the extreme
disorder limit should be eventually recovered.  Figure~\ref{f.beta}
shows how the masses $M_{f}$ and $M_{b}$ vary with the disorder
parameter $\beta$. For small values, $\beta<0.1$, both masses are
proportional to the system size, and depend weakly on $\beta$.  For
intermediate values, $0.1<\beta<\beta^*(L)$, the masses depend of
$\beta$ but still grow linearly with $L$, therefore indicating the
persistence of the weak disorder regime. However, as one goes to
larger values, $\beta>\beta^*(L)$, the curves cross over to the strong
disorder regime, where the masses show super-linear growth with system
size, $M_{t}\sim{L^{d_{t}}}$, $M_{f}\sim{L^{d_{f}}}$, and
$M_b\sim{L^{d_b}}$, but are again not dependent on $\beta$. The value
$\beta^*$ marks the transition from weak to strong disorder. One
should expect that, above a characteristic length $\xi$, the system
scales as in the weak disorder limit. Certainly, the length $\xi$
should depend on the strength of the disorder, diverging in the
extreme disorder as $\xi\sim\beta^{\frac{1}{\eta}}$. The onset of the
strong disorder regime takes place when $\xi>L$, resulting in
$\beta^*\sim{L^{\eta}}$.  The insets of Figs.~\ref{f.beta} show the
data scaled by the fractal dimensions and $\beta^*$, with the
controlling exponent $\eta=0.9\pm 0.1$. The collapse of the curves in
the transition region corroborates our analysis and reveals the
presence of a crossover between the two regimes.

In conclusion, our results show that the threshold disorder introduces
a characteristic scale $\xi$ in the system. Below this scale, the
fracture backbone displays a tortuous self-similar shape, with the
same fractal dimension of the optimum path under strong disorder
\cite{Cieplak1994,Cieplak1996} and other previously investigated
models \cite{Andrade2009,Fehr2009,Oliveira2011,Araujo2010}.  For
scales larger than $\xi$, the fractures grow linearly with system
size, consistent with the weak disorder regime. In the limit of
extreme disorder, $\xi\to\infty$, the largest fracture has a fractal
dimension of $d_f=1.86\pm{0.01}$, close to, but different from the
fractal dimension of percolation clusters \cite{Roux1988}.

We thank the Brazilian Agencies CNPq, CAPES, FUNCAP and FINEP, the
FUNCAP/CNPq Pronex grant, and the National Institute of Science and
Technology for Complex Systems in Brazil for financial support.


\begin{thebibliography}{00}
 
\bibitem{Marder2010} M. P. Marder, \textit{Condensed Matter Physics} -
  second edition (Wiley, New Jersey, 2010).

\bibitem{Kanninen1985} M. F. Kanninen and C. H. Popelar, \textit{Advanced
    Fracture Mechanics} (The Oxford Engineering Science Series, New
  York, 1985).

\bibitem{Sahimi1998} M. Sahimi, Physics Reports {\bf 306}, 213 (1998).

\bibitem{Alava2006} M. J. Alava, P. K. V. V. Nukala, and S. Zapperi,
  Adv. Phys.  \textbf{55}, 349 (2006).

\bibitem{Herrmann1988} B. Kahng, G. G. Batrouni, S. Redner, L. de Arcangelis,
  and H. J. Herrmann, Phys. Rev. B {\bf 37}, 37 (1988).

\bibitem{Malakhovsky2007} I. Malakhovsky and M. A. J. Michels, Phys. Rev.
  B {\bf 76}, 144201 (2007).

\bibitem{Hao2012} D.-P. Hao, G. Tang, H. Xia, K. Han, and Z.-P. Xun,
  J. Stat. Phys. {\bf 146}, 1203 (2012).

\bibitem{Hansen1991} A. Hansen, E. L. Hinrichsen, and S. Roux, Phys.
  Rev.  Lett. {\bf 66}, 2476 (1991); A. Hansen, E. L. Hinrichsen, and
  S. Roux, Phys. Rev. B {\bf 43}, 665 (1991).

\bibitem{Arcangelis1989} L. de Arcangelis, A. Hansen, H. J. Herrmann, and
  S. Roux, Phys. Rev. B {\bf 40}, 877 (1989).

\bibitem{Gilabert1987} A. Gilabert, C. Vanneste, D. Sornette, and E.
  Guyon, J. Physique {\bf 48}, 763 (1987).

\bibitem{Zapperi2008} M. J. Alava, P. K. V. V.  Nukala, and S.
  Zapperi, Int. J.  Fract. {\bf 154}, 51 (2008); P. K. V. V. Nukala,
  S.  Zapperi, M. J. Alava, and S. \v{S}imunovi\'{c}, Int. J.  Fract.
  {\bf 154}, 119 (2008); C. Manzato, A. Shekhawat, P. K. V. V. Nukala,
  M. J. Alava, J. P.  Sethna, and S. Zapperi, Phys. Rev. Lett.
  \textbf{108}, 065504 (2012).

\bibitem{Arcangelis1985} L. de Arcangelis, S. Redner, and H. J.
  Herrmann, J. Physique Lett. {\bf 46}, L-585 (1985).

\bibitem{footnote1} In the case where all the conductances are the
  same, some symmetric conditions may happen, where a fraction of the
  backbone lies on a Wheatstone bridge connecting sites with the same
  potential. Our geometrical model does not account for this
  possibility. However, we performed tests introducing disorder in the
  conductance, apart from the disorder in the thresholds, avoiding
  this effect, and obtained similar results. In fact, no qualitative
  difference is expected, as long as the distribution of conductance
  is bounded, not ranging over many order of magnitudes.

\bibitem{HSL} The linear equation system was solved through the HSL
  library, a collection of FORTRAN codes for large-scale scientific
  computation. See http://www.hsl.rl.ac.uk/.

\bibitem{Andrade2009} J. S. Andrade, E. A. Oliveira, A. A. Moreira,
  and H. J. Herrmann, Phys. Rev. Lett. {\bf 103}, 225503 (2009).

\bibitem{Cieplak1994} M. Cieplak, A. Maritan, and J. R. Banavar, Phys.
  Rev. Lett. {\bf 72}, 2320 (1994).

\bibitem{Fehr2009} E. Fehr, J. S. Andrade, S. D. da Cunha, L. R. da
  Silva, H. J. Herrmann, D. Kadau, C. F. Moukarzel, and E. A.
  Oliveira, J. Stat. Mech. P09007 (2009).

\bibitem{Oliveira2011} E. A. Oliveira, K. J. Schrenk, N. A. M.
  Ara\'ujo, H. J. Herrmann, and J. S. Andrade, Phys. Rev. E {\bf 83},
  046113 (2011).

\bibitem{Araujo2010} N. A. M. Ara\'ujo and H. J. Herrmann, Phys. Rev.
  Lett. {\bf 105}, 035701 (2010).

\bibitem{Cieplak1996} M. Cieplak, A. Maritan, and J. R.  Banavar,
  Phys.  Rev. Lett. {\bf 76}, 3754 (1996).

\bibitem{Stauffer1992} D. Stauffer and A. Aharony, \textit{Introduction to
    Percolation Theory} (Taylor \& Francis, London, 1992).

\bibitem{Roux1988} S. Roux, A. Hansen, H. Herrmann, and E. Guyon, J. Stat.
  Phys. {\bf 52}, 237 (1988).

\bibitem{Porto1997} M. Porto, S. Havlin, S. Schwarzer, and A. Bunde, Phys.
  Rev. Lett. {\bf 79}, 4060 (1997); M. Porto, N. Schwartz, S.  Havlin,
  and A. Bunde, Phys.  Rev. E {\bf 60}, R2448 (1999).

\bibitem{Schrenk2012} K. J. Schrenk, N. A. M. Ara\'ujo, J. S. Andrade, and
  H. J. Herrmann, Sci. Rep. {\bf 2}, 348 (2012).

\bibitem{Wilkinson1983} D. Wilkinson and J. F. Willemsen, J. Phys. A {\bf
    16}, 3365 (1983).

\bibitem{Roux89} S. Roux and E. Guyon, J. Phys. A {\bf 22}, 3693 (1989).

\bibitem{Sheppard99} A. P. Sheppard, M. A. Knackstedt, W. V. Pinczewski, and
M. Sahimi, J. Phys. A {\bf 32}, L521 (1999).

\end{thebibliography}
\end{document}